\documentstyle[pre,aps]{revtex}

\newcommand{\be}{\begin{equation}}
\newcommand{\ee}{\end{equation}}
\newcommand{\prt}{\partial}
\newcommand{\ep}{\varepsilon}

\begin{document}

\draft

\title{Ion--beam induced current in high--resistance materials}

\author{V.I. Yukalov$^{1,2}$ and E.P. Yukalova$^{1,3}$}
\address{$^1$Instituto de Fisica de S\~ao Carlos, Universidade de S\~ao Paulo \\
Caixa Postal 369, S\~ao Carlos, S\~ao Paulo 13560--970, Brazil}
\address{$^2$ Bogolubov Laboratory of Theoretical Physics \\
Joint Institute for Nuclear Research, Dubna 141980, Russia}
\address{$^3$  Department of Computational Physics \\
Laboratory of Computing Techniques and Automation \\
Joint Institute for Nuclear Research, Dubna 141980, Russia}

\maketitle

\begin{abstract}

The peculiarities of electric current in high--resistance materials, such 
as semiconductors or semimetals, irradiated by ion beams are considered. 
It is shown that after ion--beam irradiation an unusual electric current may 
arise directed against the applied voltage. Such a {\it negative current} 
is a transient effect appearing at the initial stage of the process. The 
possibility of using this effect for studying the characteristics of 
irradiated materials is discussed. A new method for defining the mean 
projected range of ions is suggested.

\end{abstract}

\vskip 1.5cm


\section{Introduction}

The irradiation of semiconductors and insulators by ion beams is a tool that
is frequently used for studying or modifying the electric properties of these
high-resistance materials (see e.g. [1,2] and references therein). As s result
of ion irradiation, heavily doped charged layers can be formed in a sample [3],
with the density of ions in the layer reaching $10^{20}$cm$^{-3}$. The formation
of strongly nonuniform charge distributions is the standard situation in the
ion-beam irradiation, while for achieving a uniform volume concentration of
implanted ions special tricks are required [4].

Under the influence of ion irradiation, electric properties of high-resistance
materials can be essentially changed [1,2,5], with additional peculiarities
resulting from a nonuniform distribution of charge carriers [6--9].
High-resistance materials mean, as usual, the materials with a very poor
concentration of carriers, such as semiconductors, semimetals, or insulators
that, being irradiated, acquire conducting properties [1,2,5]. The electron
transport in such materials can be well described in the semiclassical
drift-diffusion approach [10,11]. The existence of the implantation induced
damages can be effectively taken into account by means of characteristic
parameters entering the drift-diffusion equations [10,11], for instance, by
specifying the resistivity or mobility [1,2].

When an energetic ion strikes a solid surface, there is a probability of
electron capture resulting in the neutralization of a part of implanted ions
that become electrically inactive. However, the neutralized ions can easily be
activated again by irradiating the material with laser beams [12--15].

In the present paper, we study the electric transport in high-resistance
materials with a strongly nonuniform initial distribution of charge carriers,
which can be formed by ion-beam irradiation. As follows from the above
discussion, it is always possible to prepare such conditions, e.g. employing
laser activation [12--15], that the injected ions could be the principal
charge carriers. For concreteness, we consider below positive ions, though
this assumption is not principal and negative ions could be treated in the
same footing.

\section{Peculiarities of Electric Transport}

The transport properties of high-resistance materials, such as extrinsic
semiconductors, are usually described by the drift-diffusion approach [10,11]
consisting of the continuity and Poisson equations, respectively,
\be
\label{1}
\frac{\prt\rho}{\prt t} +\vec\nabla\cdot\vec j + \frac{\rho}{\tau} = 0 \; ,
\qquad \ep\vec\nabla\cdot\vec E = 4\pi\rho \; ,
\ee
where $\rho=\rho(\vec r,t)$ is the charge density of carriers;
$\vec j=j(\vec r,t)$ is the electric current density; $\tau$ is the relaxation
time; and $\ep$ is the dielectric permittivity. The total current density is
\be
\label{2}
\vec j_{tot} = \vec j + \frac{\ep}{4\pi}\; \frac{\prt\vec E}{\prt t} \; ,
\qquad \vec j = \mu\rho\vec E - D \vec\nabla \rho \; ,
\ee
where $\mu$ is the mobility of carriers and $D=(\mu/\ep)k_BT$ is the diffusion
coefficient. The average current through the considered sample is given by the
integral
\be
\label{3}
\vec J(t) =\frac{1}{V}\; \int\; \vec j_{tot}(\vec r,t)\; d\vec r
\ee
over the sample volume $V$.

Let us consider a plane sample of the thickness $L$ and area $A$, which is 
biased with an external constant voltage $V_0>0$. Because of the plane
symmetry, the three-dimensional picture will be reduced everywhere below
to the one-dimensional description. For what follows, it is convenient to
simplify the notation passing to dimensionless quantities. The return to
dimensional quantities can be easily done as follows. We keep in mind that
the coordinate $x$ is measured in units of the thickness $L$ and the time
$t$, in units of the transit time $\tau_0$, so that for returning to the
corresponding dimensional variables one has to make the substitution
$$
x \rightarrow \; \frac{x}{L} \; , \qquad
t \rightarrow \; \frac{t}{\tau_0} \; , \qquad
\tau_0 \equiv \; \frac{L^2}{\mu V_0} \; .
$$
For other physical quantities the return to dimensional units is done by
accomplishing the following transitions: for the diffusion coefficient,
$$
D\rightarrow \; \frac{D}{D_0} \; , \qquad D_0\equiv \mu V_0\; ,
$$
for the electric field,
$$
E\rightarrow \; \frac{E}{E_0} \; ,\qquad E_0 \equiv\; \frac{V_0}{L} \; ,
$$
for the total accumulated charge,
$$
Q \rightarrow \; \frac{Q}{Q_0} \; ,\qquad
Q_0\equiv \varepsilon A E_0 = \; \frac{\varepsilon AV_0}{L} \; ,
$$
for the charge density,
$$
\rho \rightarrow \; \frac{\rho}{\rho_0} \; , \qquad
\rho_0\equiv \;\frac{Q_0}{AL} = \; \frac{\varepsilon V_0}{L^2} \; ,
$$
and for the electric current,
$$
j \rightarrow \; \frac{j}{j_0} \; , \qquad
j_0\equiv\; \frac{Q_0}{A\tau_0} = \; \frac{\varepsilon V_0}{L\tau_0} \; ,
$$
with the same transformation for the average current (3),
$J\rightarrow J/j_0$.

Employing the dimensionless notation, for the plane case considered, we 
have from equations (1)
\be
\label{4}
\frac{\prt\rho}{\prt t} +\frac{\prt}{\prt x} \; (\rho E) - D\; 
\frac{\prt^2\rho}{\prt x^2} + \frac{\rho}{\tau} = 0 \; , \qquad
\frac{\prt E}{\prt x} = 4\pi\rho \; ,
\ee
where the space and time variables are such that
$$
0 <\; x \; < 1\; , \qquad t > \; 0 \; .
$$
The total current density (2) reads
\be
\label{5}
j_{tot} = \rho E - D\; \frac{\prt\rho}{\prt x} + \frac{1}{4\pi}\;
\frac{\prt E}{\prt t} \; .
\ee
The condition that the sample is biased with a constant external voltage 
now writes
\be
\label{6}
\int_0^1 E(x,t) \; dx = 1 \; .
\ee
Note that the parameters $\ep$ and $\mu$ do not appear in Eqs. (4) because of
the usage of the dimensionless units.

An initial condition to the continuity equation is defined by the
distribution of ions after the irradiation process. The distribution 
of implanted species can be modelled by the Gaussian form
\be
\label{7}
\rho(x,0) = \frac{Q}{Z} \; \exp\left\{ - \frac{(x-a)^2}{2b} \right\} \; ,
\ee
in which $0<a<1$ and
$$
Q = \int_0^1 \rho(x,0) \; dx \; , \qquad
Z =\int_0^1 \exp\left\{ - \frac{(x-a)^2}{2b}\right\} \; dx \; .
$$
The distribution centre, $a$, is close to the mean projected range of ions,
although may be not exactly coinciding with it.

Our aim is to study the behaviour of the electric current through the sample,
\be
\label{8}
J(t) = \int_0^1 j_{tot} (x,t) \; dx \; ,
\ee
as a function of time, when the initial distribution of carriers is given 
by the form (7). In particular, we shall show that the nonuniformity of 
the initial distribution may lead to quite unexpected behaviour of 
the current (8) when it turns against the applied voltage becoming negative,
This will be shown to be the result of the solution of the transport equations
(4).

First of all, let us emphasize that the occurrence of the negative current,
if any, can be due only to a nonuniform distribution of carriers.
Really, the current (8), with the current density (5), can be written as
\be
\label{9}
J(t) = \int_0^1 \rho(x,t) E(x,t)\; dx + 
D\left [ \rho(0,t) - \rho(1,t)\right ] \; .
\ee
If $\rho(x,t)$ does not depend on $x$, then Eq. (9), with the help of 
the voltage condition (6), immediately shows that the current is 
positive since we are dealing with a positive charge density $\rho$. As 
far as the diffusion and relaxation terms in the continuity equation 
tend to smooth the initial nonuniform distribution of carriers, this 
tells us that the negative current can happen only at the initial stage 
of the process while the charge density $\rho(x,t)$ is yet essentially 
nonuniform. The initial stage means that $t\ll 1$, when the influence of 
the diffusion and relaxation terms is yet negligible. In other words, the 
favouring conditions for the occurrence of the negative electric current 
are $D\ll 1$ and $\tau\gg 1$. It is worth recalling here that, according
to the discussion given in the Introduction, the influence of the
implantation-induced damages is assumed to be taken into account by the
corresponding values of parameters in the transport equations and that the
implanted ions are assumed to be prepared as electrically active, which can
be done by means of laser irradiation [12--15].

To demonstrate explicitly that the negative current really can occur, let 
us analyse the case of a very narrow layer of injected ions, such that
the standard deviation $b$, also called the straggling [1], is small,
$b\ll 1$. Then distribution (7) is close to
\be
\label{10}
\rho(x,0) = Q\delta(x-a) \; .
\ee
In the beginning of the process, when $t\approx 0$, we have from Eqs. 
(9) and (10)
$$
J(0) = QE(a,0) \; .
$$
Integrating the second equation in Eq. (4) yields
$$
E(a,0) = 1 + 4\pi Q \left ( a - \frac{1}{2}\right ) \; .
$$ 
With these conditions, the current (8) becomes negative provided that
\be
\label{11}
4\pi Q\left ( \frac{1}{2} - a \right ) > \; 1 \; .
\ee
From here, the inequality
\be
\label{12}
a \; < \frac{1}{2} - \frac{1}{4\pi Q} 
\ee
follows for the location of the ion layer. Since this location is inside the 
sample, we also have the inequality
\be
\label{13}
Q > \; \frac{1}{2\pi} \qquad (0\; < a \; < 1)
\ee
for the initial charge.

The above analysis demonstrates that the negative current really can 
occur provided some special conditions, as (12) and (13), hold true.
To our knowledge, there have been no experimental measurements demonstrating
the appearance of such a negative current. Therefore the picture we describe
here is a theoretical prediction of a novel effect. The occurrence of this
negative current is very sensitive to the characteristics of the irradiated
sample as well as to the initial nonuniform distribution of carriers, which
suggests the possibility of using this effect for studying the mentioned
properties. For instance, the mean projected range of irradiating ions could
be measured in this way. This can be realized as follows. Assume that for an
ion-irradiated material we observed the occurrence of the negative current
$J(0)$ at initial time. Let us compare the observed value $J(0)$ with formula
(9) where we have to substitute the distribution (7). The peak of the latter
is close to the mean projected range of the irradiating ions. With the given
$\rho(x,0)$, the right-hand side of expression (9) can be explicitly 
calculated. This is because the electric field satisfying Eq. (4), with 
the voltage condition (6), can be presented as the functional
$$
E(x,t) = 1 + 4\pi \left [ Q(x,t) - \int_0^1 Q(x,t)\; dx \right ]
$$
of the density $\rho(x,t)$ entering through
$$
Q(x,t) =  \int_0^x \rho(x',t)\; dx' \; .
$$
Hence, for a given $\rho(x,0)$, the electric field $E(x,0)$ is uniquely 
defined by the above functional. Thus, equating the calculated $J(0)$ 
from formula (9) with the corresponding measured value, we obtain an 
equation for the dimensionless mean projected range $a$. For example, in the 
case of a narrow ion layer, we find
$$
a = \frac{1}{2} - \frac{1}{4\pi Q} + \frac{J(0)}{4\pi Q^2}\; .
$$
Returning to dimensional units, for the mean projected range $\lambda =aL$
we obtain
\be
\label{14}
\lambda = \frac{L}{2} - \frac{\ep A V_0}{4\pi Q} +
\frac{\ep A^2L^2}{4\pi\mu Q^2} \; J(0) .
\ee
This formula directly relates the mean projected range of ions, $\lambda$,
with the known parameters of the system and the measured initial current 
$J(0)$. In deriving Eq. (14), we have only used the fact that the initial 
distribution of ions is narrow, $b\ll 1$, while the current $J(0)$ can, 
in general, be of any sign. The advantage of using the effect of the 
negative electric current is that the latter always requires a relatively 
narrow initial distribution. Therefore, as soon as we observe a negative 
current $J(0)$, we can employ formula (14) as a realistic approximation 
for the ion mean projected range. 

When the initial ion layer is not narrow and the current $J(0)$ is not 
necessary negative, so that the dependence of the formulas on the standard
deviation $b$ becomes important, then two situations can happen. One is
when $b$ is known from other experiments. Then, in order to find the mean 
projected range, one should proceed exactly as is described above equating
the measured and calculated currents $J(0)$. If the straggling $b$ is not 
known, it can be found in the following way. Assume that at the initial time
$t=0$ there appears the negative current. Since this is a transient 
effect, existing only in a finite time interval $0\leq t\leq t_0$, there 
is some time $t=t_0$ when the current changes its sign becoming normal, 
i.e. positive, for $t>t_0$. At the same time $t=t_0$, the current then is 
zero, $J(t_0)=0$. Therefore, the latter equation, with the given 
experimentally measured $t_0$, may serve as an additional condition 
defining the straggling $b$.

It may also happen that the injected ion layer is narrow but located 
close to the surface, so that $a\sim b$. Then formula (14) can be 
corrected by taking into account the second term in Eq. (9), which gives 
for the mean projected range
\be
\label{15}
\lambda = \frac{L}{2} - \frac{\ep A}{4\pi Q} \left [ V_0 +
\frac{AL}{\mu Q} \left ( D\Delta\rho - LJ\right )\right ] \; ,
\ee
where
$$
J\equiv J(0) \; , \qquad \Delta\rho \equiv \rho(0,0) - \rho(1,0) \; .
$$

To find the electric current (8) for arbitrary times, we need to solve 
equations (4), with the voltage condition (6) and the initial 
condition (7). Those should also be complimented by boundary conditions 
which can be taken in the Neumann form [10]. We have accomplished such 
numerical calculations whose results are presented in the figures, where 
all values are given in dimensionless units, and the conditions $D\ll 1$ and
$\tau\gg 1$, favouring the occurrence of the negative current are assumed.
The figures correspond to the setup explained in detail in the Introduction
and specified in the present section.

Fig. 1 shows the dependence of the electric current $J(t)$ on time for
the varying location $a$ of the initial distribution peak of ions. 
Formulas (14) and, respectively, (15) for the mean projected range are valid 
with a very good accuracy, with an error being less than $1\%$. In order 
to distinguish the lines, they are numbered.

Fig. 2 presents the electric current as a function of time for different 
widths of the initial distribution of ions. Formulas (14) and (15) are 
again very accurate for the negative current $J(0)$. But when the current 
$J(0)$ is positive and at the same time, the initial distribution is not 
narrow, then Eqs. (14) and (15) are not valid, as it should be.

Fig. 3 demonstrates the time dependence of the electric current for 
different values of the charge $Q$. Since $b\ll 1$, formulas (14) and (15)
work perfectly for both cases, for the negative current, as in Fig. 3, as
well as for positive current, as in Fig. 4; the errors being not more than
around $1\%$. The current in Fig. 4 becomes positive because condition (13)
does not hold. Increasing the accumulated charge $Q$, so that condition (13) 
becomes true, the negative current appears provided that condition (12) 
is satisfied, as is clearly illustrated in Fig. 5.

\section{Discussion}

We have shown that in high resistance materials irradiated by ion beams an
unusual transient effect of the negative electric current can occur. This
effect can be used for studying the characteristics of irradiated materials
as well as those of irradiating ions. For instance, the mean projected range
of ions can be measured whose value is well approximated by formula (15).
This new way of measuring the ion mean projected range is, certainly, not the
sole possible, but it can provide additional information being complimented
by other experimental methods. Throughout the paper, we have been talking
about ions. But, as is clear, the same consideration concerns any kind of
charged particles, say electrons, positrons, or muons.

It is worth stressing that here we have advanced a theoretical proposal
for observing and employing a novel effect. The general prerequisites for
realizing the latter look, as follows from the above discussions, achievable.
Moreover, there exist so many types of various high-resistance materials,
such as semiconductors, semimetals, and insulators, there are so many ways
of fabricating such materials with specially designed characteristics, and
also there are so many methods of varying the properties of these materials
by involving additional external sources, as electromagnetic fields or laser
beams, that the feasibility of preparing the desired conditions for realizing
the suggested effect looks quite realistic. Even if this realization is
difficult today, it might be accomplished in future. The history shows that
practically any effect that is possible in principle sooner or later becomes
realizable in experiment.

Although our main aim in this paper has been to advance a theoretical
proposal for the {\it principal possibility} of observing a novel effect,
we have also paid attention to the feasibility of realizing it
experimentally. In addition to the above discussions, we would like to touch
several other important points that could substantiate the feasibility of
observing the considered effect.

First, the space charge due to the accumulated ions will act against further
implantation, and the ion doses will certainly be limited. At the same time,
to observe the effect of negative electric current, one needs to reach some
threshold value for the accumulated charge, as in the inequality (13). In
order to understand how the latter condition can be satisfied, it is necessary
to rewrite it in the corresponding dimensional units introduced in the
beginning of Sec. 2, in this case, in the units of $Q_0$. This gives us the
condition
$$
2\pi\ep A\;\frac{V_0}{L}\; Q > 1\; .
$$
From here, it is evident that this condition can always be satisfied for any
given charge $Q$, which can be achieved by increasing either the sample
area $A$ or the applied voltage $V_0$. Of course, for a given sample, one has
to increase the voltage $V_0$.

The implanted ions do not form a homogeneous layer and their distribution may
be not exactly Gaussian. In this paper we consider the Gaussian distribution
(7) which usually describes well the profile of implanted ions. However this form
is not compulsory for realizing the negative-current effect. The latter
persists as well for other distributions, for which one needs only that the
straggling $b$ be less than the projected range $a$.

For the realization of the effect, it is also not necessary that the ion
distribution be centered at the mean projected range of ions. For simplicity,
we called the distribution center $a$ the projected range. However, if these
two do not coincide, all consideration remains valid with merely slightly
changing the terminology, so that $a$ is to be called the distribution center.

To measure currents through the sample, one has to take into account that
near contacts there often occurs the space charge build-up caused by electron
or hole injection from contacts. The current due to the carriers injected
from contacts can mask the current produced by ions. How to ascertain that
the considered effect is caused by the implanted ions? This problem can be
easily resolved by measuring the current through the sample before ion
irradiation. Then one may study the influence of contacts on creating
electric currents. In this way, one can always separate the influence of
contacts from physical effects resulting from ion irradiation.

In conclusion, we do not see principal difficulties for realizing the
negative-current effect. And, as follows from the above discussions,
different technical problems seem to be resolvable. Being realized, this
effect will make it possible to have an additional tool for studying the
transport properties of high-resistance materials as well as to measure
characteristics related to irradiated ions.

\vskip 5mm

{\bf Acknowledgement}

\vskip 2mm

We appreciate financial support from the S\~ao Paulo State Research
Foundation.

\newpage

\begin{center}
{\bf Figure Captions}
\end{center}

{\bf Fig. 1.} 

\vskip 2mm

Electric current through the sample as a function of time 
for the parameters $Q=1,\; b=0.05$,
and different initial locations of the ion layer: $a=0.1$ (curve 1, solid 
line), $a=0.25$ (curve 2, long--dashed 
line), $a=0.5$ (curve 3, short--dashed line), and  $a=0.75$ (curve 4, dotted 
line).

\vskip 0.5cm

{\bf Fig. 2.} 

\vskip 2mm

The dependence of the electric current on time for  $Q=0.5,\; a=0.1$, 
and varying widths of the initial ion distribution: $b=0.05$ (curve 1, solid 
line), $b=0.1$ (curve 2, long--dashed line), and $b=0.5$ (curve 3, 
short--dashed line).

\vskip 0.5cm

{\bf Fig. 3.} 

\vskip 2mm

Electric current as a function of time for $a=0.25,\; b=0.05$ and
different charges: $Q=3$ (curve 1, solid line), $Q=1$ 
(curve 2, long--dashed line), and $Q=0.5$ (curve 3, short--dashed line).

\vskip 0.5cm

{\bf Fig. 4.} 

\vskip 2mm

The time dependence of the electric current for  $Q=0.1,\; b=0.05$, 
and two different initial locations of the ion distribution: $a=0.1$ 
(curve 1, solid line) and $a=0.25$ (curve 2, long--dashed line).

\vskip 0.5cm

{\bf Fig. 5.} 

\vskip 2mm

Electric current vs. time for $Q=0.5,\; b=0.1$, and varying initial 
locations of the peak of the ion distribution: $a=0.1$ (curve 1, solid line),
$a=0.25$ (curve 2, long--dashed line), $a=0.5$ (curve 3, short--dashed 
line), and $a=0.75$ (curve 4, dotted line).

\end{document}